\documentclass[twocolumn,showpacs,preprintnumbers,draftclsnofoot]{revtex4}
\usepackage{amsmath}
\usepackage{amssymb}
\usepackage{graphicx}
\usepackage{float}
\usepackage{placeins}
\usepackage{dcolumn}
\usepackage{bm}
\usepackage{amssymb}
\usepackage{graphicx}
\usepackage{dcolumn}
\usepackage{bm}
\usepackage{hyperref}
\begin{document}

\title{ Optical Amplification with Large Goos-H$\ddot{a}$nchen Shift Driven by Non-Hermitian Bilayer Meta-Grating Under Lasing State }
\author{ Ma Luo \footnote{Corresponding author:luoma@gpnu.edu.cn} and Xueyi Zhang }
\affiliation{School of Physics and Optoelectronic Engineering, Guangdong Polytechnic Normal University, Guangzhou 510665, China  }

\begin{abstract}

The coexistence of gain and loss in parity-time symmetric bilayer meta-gratings enables the transformation of bound states in the continuum into a ring of lasing threshold modes. Under lasing state, gain saturation drives the contraction of this mode ring into a single discrete lasing point. Additionally incident weak optical field with frequency and incident angle nearly matched to the lasing mode can obtain amplification accompanied by prominent Goos-H$\ddot{a}$nchen shifts. Specifically, active meta-grating with lasing mode at the $\Gamma$ point induce sizable Goos-H$\ddot{a}$nchen shifts. In contrast, active meta-grating with lasing mode off the $\Gamma$ point induce large Goos-H$\ddot{a}$nchen shifts that are proportional to the quality factor of the effective resonant modes under lasing state.  In finite meta-grating, finite-size effects exert distinct influences on the lasing state as well as the Goos-H$\ddot{a}$nchen shifts of the additionally incident optical field: the lasing characteristics and Goos-H$\ddot{a}$nchen shifts of $\Gamma$-point lasing modes remain nearly unaffected, whereas those of off-$\Gamma$ lasing modes are significantly modulated by finite-size perturbations.

\end{abstract}

\maketitle

\section{Introduction}

The Goos-H$\ddot{a}$nchen (GH) shift describes the lateral displacement of a finite-width optical beam away from its geometric-optical propagation trajectory \cite{firstGH47}. According to the stationary-phase theory, this spatial shift is governed by the angular derivative of the reflection or transmission phase \cite{stationaryPt48}. The beam-steering capability enabled by the GH shift has advanced the development of diverse photonic devices, including optical sensors \cite{GHapplSensing1,GHapplSensing2,GHapplSensing3}, optical data storage \cite{GHapplStore1,GHapplStore2,GHapplStore3,GHapplStore4}, wavelength-division demultiplexers \cite{GHapplMulti}, optical switches \cite{GHapplSwitch}, and polarization beam splitters \cite{GHapplBeam}. Since these practical applications fundamentally require a large GH shift magnitude, extensive efforts have been devoted to exploring various photonic resonant structures for GH shift enhancement. Representative configurations include dielectric interfaces supporting the Brewster effect \cite{Brewster1,Brewster2,Brewster3}, metallic thin films featuring surface plasmon polaritons \cite{sppGH1,sppGH2,sppGH3,sppGH4,sppGH5}, Fabry-P$\acute{e}$rot resonant cavities \cite{fpcavityGH1,fpcavityGH2,fpcavityGH3,fpcavityGH4}, multilayer thin films sustaining Bloch surface waves \cite{blochGH1,blochGH2}, metal-dielectric multilayer interfaces with Tamm plasmon polaritons \cite{tamnGH1,tamnGH2,tamnGH3}, dielectric metagratings integrated with zone-folding quasi-bound states in the continuum (quasi-BICs) \cite{bicGH1,bicGH2,bicGH3}, and magneto-optical metagratings possessing $\Gamma$-point zone-folding quasi-BICs \cite{normalGH1,normalGH2,normalGH3,normalGH4,normalGH5}. Recent investigations further reveal that metagratings with unidirectional guided resonances (UGRs) can yield prominent GH shifts even without sharp reflection resonances \cite{ugrghshift1,ugrghshift2,ugrghshift3}, allowing narrow-width incident Gaussian beams to produce fully transmitted beams with remarkable GH displacement.

Non-Hermitian photonics represents a thriving research frontier, where optical gain is introduced to compensate intrinsic material loss and flexibly modulate the decay rate of resonant modes \cite{nonHermian0,nonHermian1,nonHermian2,nonHermian3}. In parity-time ($\mathcal{PT}$)-symmetric systems, the balanced distribution of gain and loss enables the coalescence of eigenstates at exceptional points (EPs) with degenerate eigenvalues \cite{nonHermian3,EPpoint2}. Periodic $\mathcal{PT}$-symmetric photonic structures have demonstrated numerous exotic phenomena, including asymmetric diffraction \cite{PTgrating1,PTgrating2}, phase-free transmission \cite{PTgrating3}, and spectral singularities \cite{PTgrating4}. Conventional GH shift enhancement strategies predominantly depend on the excitation of high-Q resonant modes with ultralow decay rates. Nevertheless, inherent material loss fundamentally limits the achievable Q factor, thereby restricting the maximum enhancement of GH shifts. To address this bottleneck, gain-loss compensation schemes have been adopted to elevate the effective Q factor and further boost GH shift magnitudes \cite{PTforGH1,PTforGH2}. For photonic systems supporting bound states in the continuum (BICs), the introduction of $\mathcal{PT}$-symmetric gain and loss transforms the discrete BIC point in momentum space into a continuous ring of lasing threshold modes (LTMs) \cite{lasingTM1,lasingTM2}. Such LTMs feature real resonant frequencies and nonvanishing radiative loss. Notably, several discrete positions on the LTM ring possess zero radiative loss, which constitute a novel class of BICs termed $\mathcal{PT}$-symmetric BICs ($\mathcal{PT}$-BICs) \cite{lasingTM1}. Moreover, $\mathcal{PT}$-symmetric metagratings can support unidirectional lasing under tailored optical excitation conditions \cite{lasingTM3}.

\begin{figure}[tbp]
\scalebox{0.2}{\includegraphics{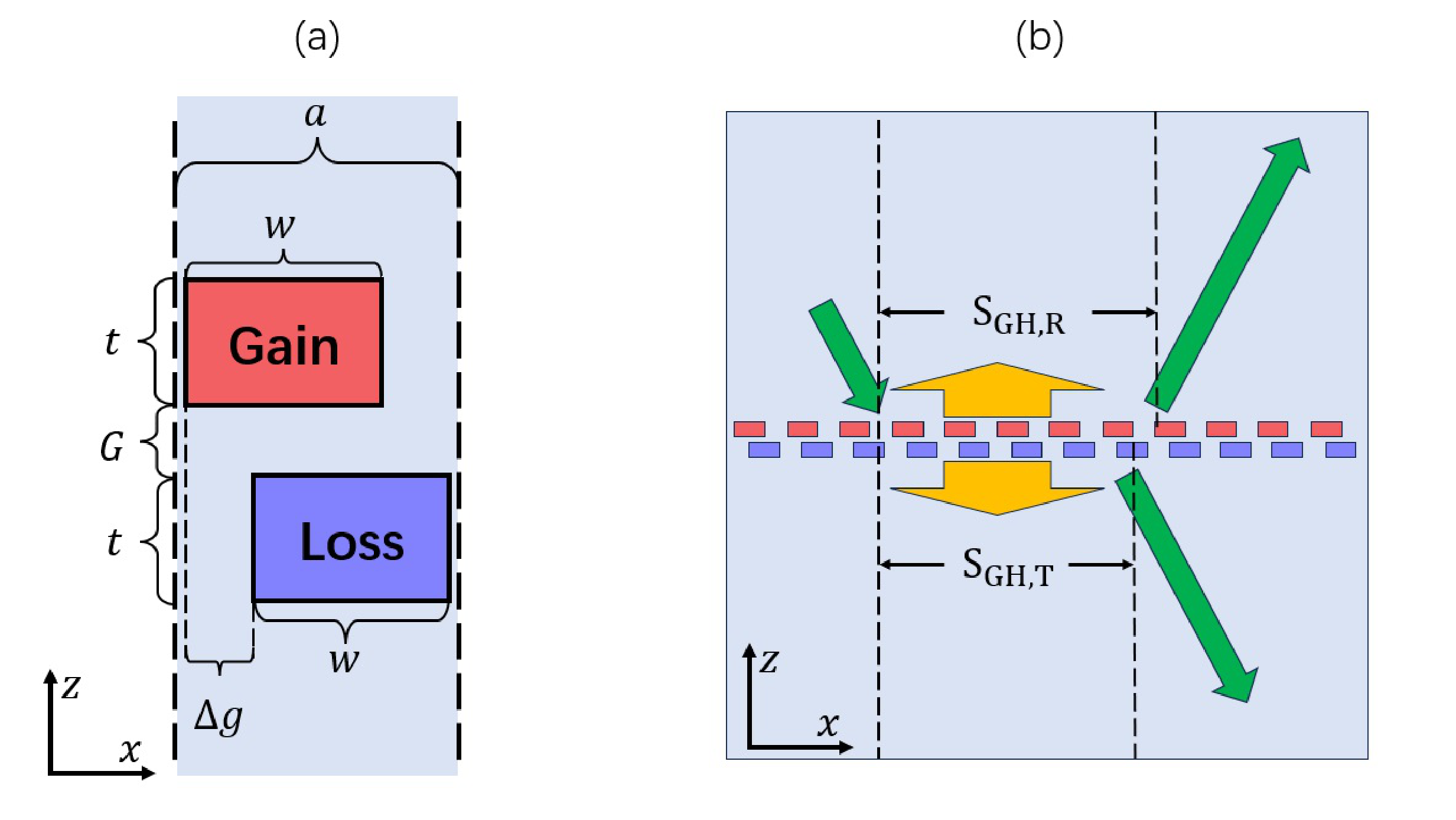}}
\caption{ (a) Schematic of a single period of the $\mathcal{PT}$-symmetric bilayer meta-grating, which is uniform along the $y$-axis and periodically modulated along the $x$-axis with a lattice period $a$. The thickness and width of both the top and bottom grating strips are denoted as $t$ and $w$, respectively, while the interlayer gap and lateral offset between the two grating layers are defined as $G$ and $\Delta g$. The gain and loss regions are marked in red and blue, respectively. (b) Illustration of non-Hermitian beam shift manipulation. The orange arrows represent the steady-state lasing background field, and the green arrows denote the additionally incident weak optical beams. The additionally incident beams are amplified and generate reflected and transmitted exiting beams with prominent GH shifts. }
\label{figure_scheme}
\end{figure}

This work theoretically investigates the large GH shift and optical amplification of additionally incident weak optical beams enabled by $\mathcal{PT}$-symmetric bilayer meta-gratings under lasing state. The studied structure is a laterally offset bilayer meta-grating, with detailed geometric parameters schematically illustrated in Fig. \ref{figure_scheme}(a). The upper and lower layers serve as gain and loss media, with refractive indices defined as $n_{g}=n_{r}-i\alpha_{g}$ and $n_{l}=n_{r}+i\alpha$ ($n_{r}=3.4767$), respectively. Exact $\mathcal{PT}$ symmetry is achieved when the gain and loss coefficients satisfy $\alpha_{g}=\alpha$. The $\mathcal{PT}$-symmetric configuration transforms original BICs into closed LTM rings, allowing the meta-grating to operate slightly above the lasing threshold. In the steady state, one resonant mode with a negative imaginary frequency eigenvalue selected from within the LTM ring evolves into the dominant lasing mode. During lasing oscillation, gain saturation suppresses the effective gain coefficient, rendering $\alpha_{g}$ a spatially dependent quantity: $\tilde{\alpha}_{g}(\mathbf{r})=\alpha_{g}/(1+I_{lasing}(\mathbf{r})/I_{satur})$, where $I_{lasing}$ denotes the local intensity of the lasing mode and $I_{satur}$ represents the saturation intensity. With the saturated gain profile $\tilde{\alpha}_{g}(\mathbf{r})$, all resonant modes in the photonic band structure exhibit positive imaginary frequencies (i.e., radiative decay), while the lasing mode maintains a vanishing imaginary frequency corresponding to stable resonance. When an additional optical beam is incident on the meta-grating, the total local intensity becomes $I_{lasing}+I_{inc}$ (where $I_{inc}$ is the intensity contributed by the incident beam), which further reduces the effective gain. For sufficiently weak incidence satisfying $I_{inc}\ll I_{lasing}$, the intrinsic lasing behavior remains nearly undisturbed. The schematic figure of the physical model is indicated in Fig. \ref{figure_scheme}(b), demonstrating coexistence of the lasing mode with finite radiation and the additionally incident weak optical field that experiences amplification with GH shift. When the incident frequency and angle match with the resonant modes near to the lasing mode, the system yields strongly amplified resonant peaks in both reflectance and transmittance, accompanied by rapidly varying reflection and transmission phases. Based on the stationary-phase theory, such steep phase variations produce significantly enhanced GH shifts for quasi-plane-wave incidence. Furthermore, for practical meta-gratings with finite periodicity, boundary-induced lateral radiative loss at the grating termini substantially modulates both the lasing performance and the GH shift characteristics of the additionally incident weak optical beam.

The remainder of this paper is organized as follows. Section II elaborates on the lasing characteristics of the $\mathcal{PT}$-symmetric meta-grating and investigates its corresponding resonant band structure. In Section III, we explore the optical amplification and prominent GH shifts of reflected and transmitted beams under weak incidental optical excitation. Theoretical analysis is performed based on the stationary-phase theory, and numerical validation is conducted via Gaussian beam simulations. Finally, Section IV summarizes the key conclusions of this work.

\section{Band structure and resonant modes under lasing state}

To demonstrate the lasing behaviors and the associated band structures, we investigate two typical meta-grating configurations with distinct BICs in their Hermitian limits. Structure I supports a BIC at the $\Gamma$ point protected by central inversion symmetry, while Structure II simultaneously accommodates symmetry-protected BICs and accidental BICs. Both configurations share identical geometric parameters: $w=0.704a$ and $G=0.045a$. The remaining structural parameters are set as $t=0.355a$ and $\Delta g=0.113a$ for Structure I, and $t=0.9275a$ and $\Delta g=0$ for Structure II. The background medium possesses a refractive index of $n_{b}=1.444$, whereas the refractive indices of the gain and loss blocks follow the definitions of $n_{g}$ and $n_{l}$ provided in the previous section. In this model, only the y-oriented dipole component is non-vanishing within the gain medium, such that optical gain is exclusively applied to y-polarized electric fields. Accordingly, only the y-axis diagonal component of the refractive index tensor in the gain material carries a negative imaginary part, while the x- and z-axis components remain purely real. In contrast, all diagonal refractive index components of the lossy medium exhibit positive imaginary parts. Such anisotropic gain-loss distribution restricts lasing to TE modes, where only the field components $(H_{x},E_{y},H_{z})$ are nonzero.

The complex resonant frequency $f$ as a function of the in-plane wavevector $k_{x}$ is numerically solved using the finite element method (FEM) implemented in the commercial software COMSOL Multiphysics. The first three TE-mode bands are denoted as TE$_{1}$, TE$_{2}$, and TE$_{3}$, respectively. The computational unit cell is constructed according to the geometry illustrated in Fig. \ref{figure_scheme}(a). Bloch periodic boundary conditions with a phase factor of $e^{ik_{x}a}$ are imposed on the left and right boundaries. The top and bottom boundaries at $z=\pm z_{max}$ are terminated by perfectly matched layers (PMLs) with a thickness of $2a$. For each resonant mode, the complex amplitudes of upward and downward radiated plane waves, denoted as $c_{u}$ and $c_{d}$, are extracted via Fourier integration: $c_{u/d}=(1/a)\int_{z=\pm z_{max}}E_{y}e^{ik_{x}x}dx$. The spatial field profile $E_{y}$ is normalized such that the maximum optical intensity inside the gain block equals the saturation intensity, namely $max[I_{res}(\mathbf{r})]=I_{satur}$, where $I_{res}(\mathbf{r})=|\mathbf{P}(\mathbf{r})|$ and $\mathbf{P}$ represents the time-averaged Poynting vector derived from the field profile $E_{y}$. The saturation intensity of the gain medium is assumed to be $I_{satur}=2$ $MW/cm^{2}$. To prevent numerical artifacts arising from overlap between evanescent resonant fields and the PML boundaries, a sufficiently large domain height $z_{max}=15a$ is adopted to guarantee simulation accuracy. Conventionally, the quality (Q) factor is defined as $Q=\mathrm{Re}[f]/(2\mathrm{Im}[f])$, where $\mathrm{Re}[f]$ and $\mathrm{Im}[f]$ correspond to the real and imaginary parts of the complex resonant frequency $f$, respectively. Nevertheless, non-Hermitian photonic systems with divergent Q factors can still sustain radiative loss. Therefore, a high Q value alone is insufficient to identify BICs in non-Hermitian meta-gratings. Instead, BICs are unambiguously characterized by fully suppressed radiation amplitudes, i.e., $|c_{u/d}|\to0$.

\begin{figure}[tbp]
\scalebox{0.6}{\includegraphics{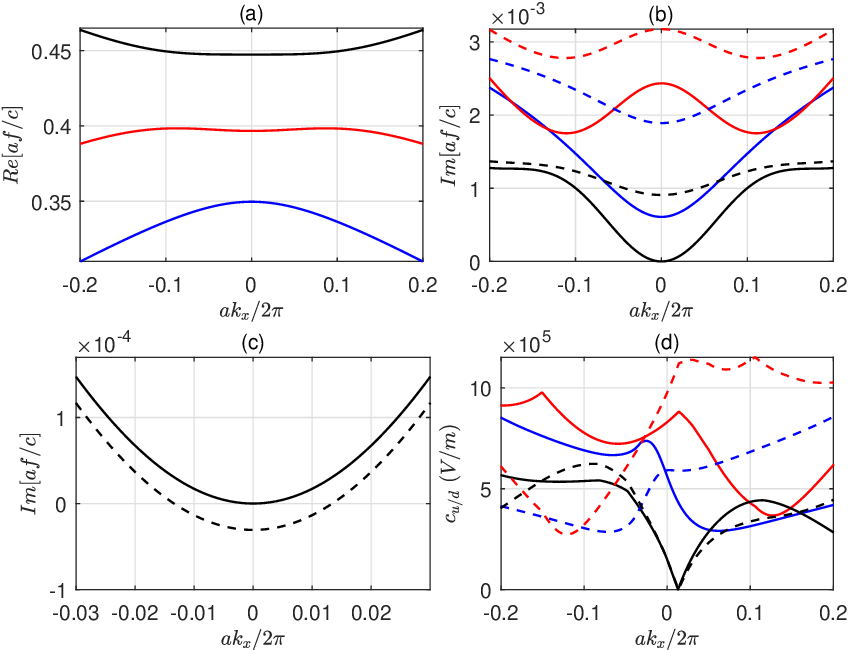}}
\caption{ The real and imaginary parts of the resonant frequency band structures for TE modes of the Structure I bilayer meta-grating at $\alpha=\alpha_{g}=0$ are depicted by solid lines in panels (a) and (b), respectively. The $\text{TE}_{1}$, $\text{TE}_{2}$, and $\text{TE}_{3}$ bands are colored in blue, red, and black, correspondingly. For the case of $\alpha=0.05$ and $\alpha_{g}=0$, the imaginary parts of the three bands are presented as dashed lines in panel (b). (c) Under fixed $\alpha=0.05$, the imaginary parts of the three bands are plotted with solid lines for $\alpha_{g}=0.04944$ and dashed lines for $\alpha_{g}=0.05$. This panel provides a zoom-in view of the spectral regime near $\mathrm{Im}[f]\approx0$. (d) For the non-Hermitian meta-grating under lasing state, the magnitudes of the radiative coefficients $c_{u/d}$ extracted from the normalized linear resonant modes in the band structure are illustrated. Here, $|c_{u}|$ and $|c_{d}|$ are represented by solid and dashed lines, with line colors consistent with the corresponding bands in panel (a).  }
\label{figure_band755}
\end{figure}

\begin{figure}[tbp]
\scalebox{0.58}{\includegraphics{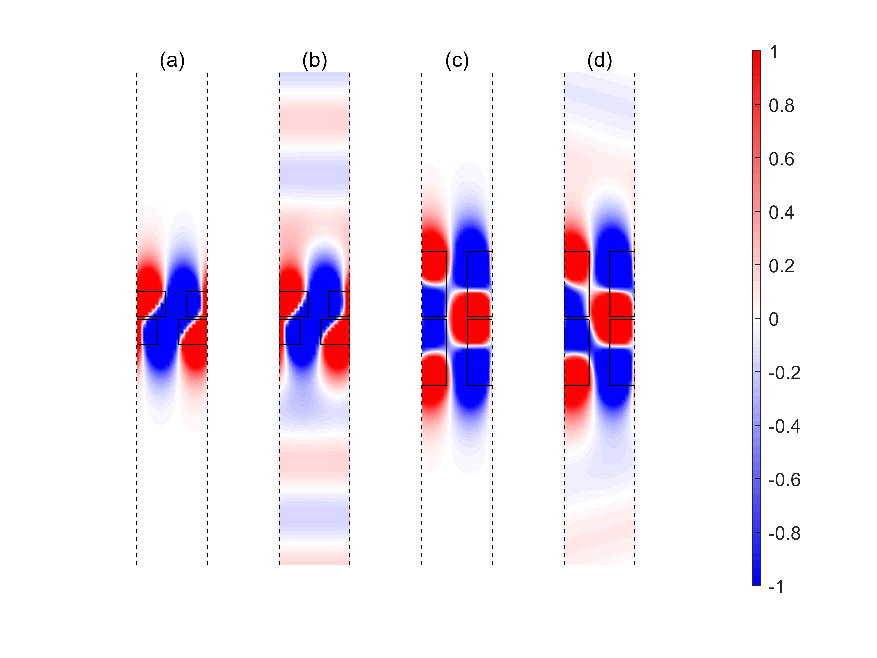}}
\caption{ Transverse electric field ($E_{y}$) distributions of the resonant modes. (a) Field profile of the BIC located at $k_{x}=0$ in the TE$_{3}$ band, corresponding to the black dispersion curve in Fig. \ref{figure_band755}(a). (b) Field profile of the LTM at $k_{x}=0$ for the case of $\alpha_{g}=0.04944$ in Fig. \ref{figure_band755}(c). (c) Field profile of the accidental BIC at $ak_{x}/2\pi=0.145$ in the TE$_{3}$ band, corresponding to the black curve in Fig. \ref{figure_band1900}(a). (d) Field profile of the LTM at $ak_{x}/2\pi=0.145$ for the case of $\alpha_{g}=0.04989$ in Fig. \ref{figure_band1900}(c). }
\label{figure_bandmode}
\end{figure}

For Structure I in the Hermitian limit ($\alpha_{g}=\alpha=0$), the real and imaginary parts of the resonant frequency bands are plotted as solid lines in Figs. \ref{figure_band755}(a) and (b), respectively. At the $\Gamma$ point, the TE$_{3}$ band exhibits a zero imaginary frequency component, corresponding to a BIC state, whose field pattern is plotted in Fig. \ref{figure_bandmode}(a). Notably, the structure breaks mirror symmetry, which indicates that this BIC is not conventional symmetry-protected BIC; instead, this BIC is topologically protected by the central inversion symmetry of the meta-grating \cite{raDongJianWen}. As $\alpha_{g}$ and $\alpha$ vary within the range from $0$ to $0.05$, the real-part band structure remains nearly unchanged, whereas the imaginary-part band structure undergoes dramatic modifications.     In practical device implementations, the loss coefficient $\alpha$ is fixed after fabrication, while the gain coefficient $\alpha_{g}$ can be dynamically tuned via pumping. In the passive state with $\alpha_{g}=0$, increasing $\alpha$ to $0.05$ elevates $Im[f]$ to  positive values, as depicted by the dashed curves in Fig. \ref{figure_band755}(b). Thus, the BIC is transferred into quasi-BICs with finite radiative loss. With activated optical pumping, $Im[f]$ of the quasi-BIC gradually decreases as $\alpha_{g}$ rises. When $\alpha_{g}$ reaches the threshold value $\alpha_{g,th}=0.04944$, $Im[f]$ of the $\Gamma$-point quasi-BIC vanishes, transforming the quasi-BIC into a LTM, as illustrated by the solid line in Fig. \ref{figure_band755}(c).  The field pattern of the LTM is plotted in Fig. \ref{figure_bandmode}(b), exhibiting finite radiation. In this case, all resonant modes are decaying except the LTM, which is non-decaying with finite radiation; accordingly, the metagrating operates at the lasing threshold \cite{lasingThreshold}. Since $\alpha_{g,th}$ is slightly lower than the loss coefficient $\alpha$, further increasing $\alpha_{g}$ to match $\alpha$ expands the discrete $\Gamma$-point LTM into a continuous LTM ring in momentum space. At $\alpha_{g}=0.05$, $Im[f]$ crosses the horizontal axis twice, as shown by the dashed curve in Fig. \ref{figure_band755}(c), which represents the one-dimensional cross-section of the full LTM ring in the two-dimensional $(k_{x},k_{y})$ momentum space. Accordingly, the $\mathcal{PT}$-symmetric meta-grating operates slightly above the lasing threshold, enabling any mode inside the LTM ring to potentially lase. Given that the $\Gamma$-point LTM with $\alpha_{g}=\alpha_{g,th}$ emerges first as $\alpha_{g}$ increases from $0$ to $0.05$, it is identified as the dominant lasing mode under steady-state operation.     During lasing oscillation, the optical gain of the top layer is suppressed by gain saturation. Owing to the slight excess of $\alpha_{g}$ above the threshold $\alpha_{g,th}$, the field distribution of the lasing mode closely resembles that of the $\Gamma$-point LTM with $\alpha_{g}=\alpha_{g,th}$. Thus, the saturated gain profile can be approximated as
\begin{equation}
\tilde{\alpha}_{g}(\mathbf{r})=\alpha_{g}/\big(1+I_{l}|\mathbf{P}(\mathbf{r})|/I_{\mathrm{satur}}\big)\label{saturGain}
\end{equation}
where $\mathbf{P}(\mathbf{r})$ denotes the Poynting vector of the $\Gamma$-point LTM with $\alpha_{g}=\alpha_{g,th}$, and $I_{l}$ is the dimensionless amplitude of the lasing mode.  To determine the stable lasing state, we numerically calculate the band structure with the saturated gain profile $\tilde{\alpha}_{g}(\mathbf{r})$ by sweeping the value of $I_{l}$. A stable lasing condition is achieved when only the dominant lasing mode possesses a purely real resonant frequency, while all other resonant modes exhibit positive $Im[f]$ corresponding to decaying modes. The optimized dimensionless amplitude is numerically given as $I_{l}=0.0212886$. Under this stable lasing state, the radiative amplitudes of all normalized TE modes are summarized in Fig. \ref{figure_band755}(d). Considering the normalization condition $max[|\mathbf{P}(\mathbf{r})|]=I_{\mathrm{satur}}$ of the field pattern, the effective radiative amplitude of the lasing mode at the $\Gamma$ point under lasing state is quantified as $\sqrt{I_{l}}|c_{u/d}|$.

Based on the temporal coupled-mode theory (TCMT) established in previous studies, the GH shift can be quantitatively evaluated as follows \cite{ugrghshift3}:
\begin{equation}
S_{GH,r(t)}\approx\lambda \frac{v_{g}}{c}\frac{\mathrm{Re}[f]}{\mathrm{Im}[f]}\label{GH_TCMT}
\end{equation}
where $\lambda$ denote the wavelength of the excitation beam in vacuum, $v_{g}$ represents the group velocity of the resonant mode, and $c$ is the speed of light in vacuum. According to this formula, a smaller $|\mathrm{Im}[f]|$ contributes to GH shift enhancement, while a reduced group velocity $v_{g}$ imposes an opposite suppression effect on the GH shift magnitude. Under lasing state, the modes surrounding the $\Gamma$-point lasing mode reside within the flat region of the TE$_{3}$ band, which features an extremely low group velocity. Such band characteristics theoretically yield a small GH shift magnitude for near-$\Gamma$ resonant modes.

\begin{figure}[tbp]
\scalebox{0.6}{\includegraphics{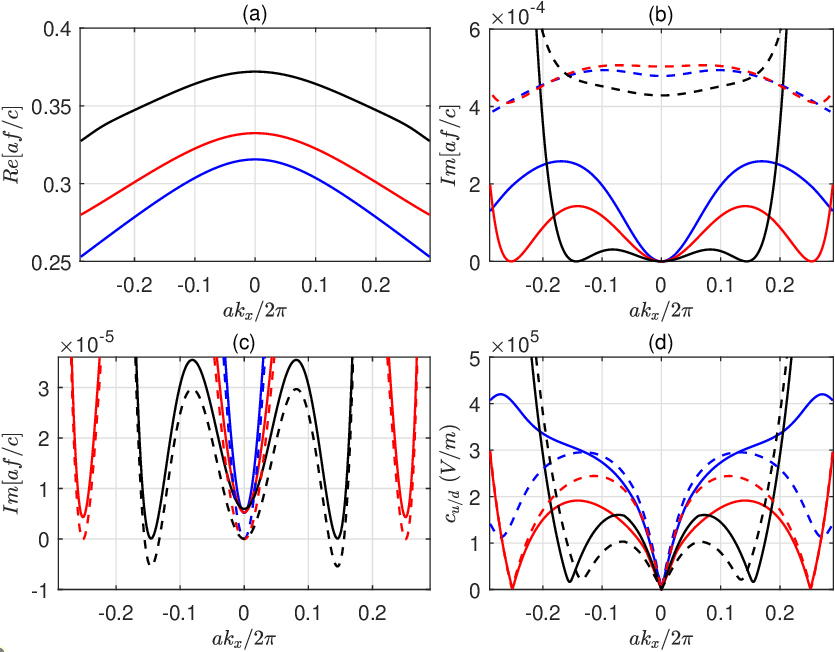}}
\caption{ The same type of plotted as those in Fig. \ref{figure_band755} except for structure II. In (c), $\alpha_{g}=0.04989$ and $0.05$ for solid and dashed lines, respectively.    }
\label{figure_band1900}
\end{figure}

For Structure II in the Hermitian limit ($\alpha_{g}=\alpha=0$), the real and imaginary parts of the resonant frequency bands are depicted by the solid lines in Figs. \ref{figure_band1900}(a) and (b), respectively. All three TE bands support symmetry-protected BICs at the $\Gamma$ point, while the TE$_{2}$ and TE$_{3}$ bands additionally host accidental BICs. The field distribution of the accidental BIC residing on the TE$_{3}$ band at $k_{x}=0.145(2\pi/a)$ is illustrated in Fig. \ref{figure_bandmode}(c). The lasing state of Structure II are analyzed following the same workflow adopted for Structure I.     With fixed loss coefficient $\alpha=0.05$ and zero gain ($\alpha_{g}=0$), $Im[f]$ of all three bands shift to positive values, as indicated by the dashed curves in Fig. \ref{figure_band1900}(b). Consequently, both symmetry-protected and accidental BICs are converted into lossy quasi-BICs. As the gain coefficient $\alpha_{g}$ rises to the threshold value $\alpha_{g,th}=0.04989$, the pair of quasi-BICs originating from the TE$_{3}$-band accidental BICs at $k_{x}=\pm0.145(2\pi/a)$ are transformed into LTMs with  $Im[f]=0$, while all other resonant modes remain decaying with $Im[f]>0$, as shown by the solid lines in Fig. \ref{figure_band1900}(c). The field pattern of the LTM at $k_{x}=0.145(2\pi/a)$ on the TE$_{3}$ band is presented in Fig. \ref{figure_bandmode}(d), which confirms the existence of finite radiation. Further increasing $\alpha_{g}$ to 0.05 establishes a strict $\mathcal{PT}$-symmetric condition for the meta-grating, which operates beyond the lasing threshold. The two discrete LTMs on the TE$_{3}$ band expand into two continuous LTM rings in momentum space, whereas the remaining quasi-BICs evolve into discrete LTM points, as visualized by the dashed curves in Fig. \ref{figure_band1900}(c). Since the paired LTMs at $k_{x}=\pm0.145(2\pi/a)$ on the TE$_{3}$ band reach the lasing threshold prior to all other modes as $\alpha_{g}$ increases from 0 to 0.05, these two modes are identified as the dominant lasing modes under steady-state operation.    Under the assumption of equal amplitude for the two paired lasing modes, the spatially dependent saturated gain profile is formulated as $\tilde{\alpha}_{g}(\mathbf{r})=\alpha_{g}/\big(1+I_{l}(|\mathbf{P}_{+}(\mathbf{r})|+|\mathbf{P}_{-}(\mathbf{r})|)/I_{\mathrm{satur}}\big)$, where $\mathbf{P}_{\pm}(\mathbf{r})$ represent the Poynting vectors of the LTMs at $k_{x}=\pm0.145(2\pi/a)$ for $\alpha_{g}=\alpha_{g,th}$. The numerically solved dimensionless lasing amplitude is given as $I_{l}=0.003742025$. The radiative amplitudes of all normalized TE resonant modes under the stable lasing state are summarized in Fig. \ref{figure_band1900}(d). Modes in the vicinity of the lasing mode exhibit asymmetric radiative losses along the upward and downward propagation directions. Such anisotropic radiation characteristics can be quantitatively evaluated via the directionality factor, defined as $\eta=(|c_{u}|^2-|c_{d}|^2)/(|c_{u}|^2+|c_{d}|^2)$. The directionality of the resonant modes with frequency lower and higher than the lasing frequency have opposite sign. The maximum magnitude of $\eta$ occurs at $k_{x}=\pm0.138(2\pi/a)$ and $k_{x}=\pm0.152(2\pi/a)$, yielding directionality values of $0.76$ and $-0.84$, respectively.

For the two proposed $\mathcal{PT}$-symmetric meta-grating structures, the emergent lasing modes originate from two distinct types of BICs, namely centrally inversion-symmetry-protected BICs and accidental BICs. Symmetry-protected BICs exhibit low sensitivity to the balanced $\mathcal{PT}$-symmetric coefficient $\alpha=\alpha_{g}$. Although such symmetry-protected BICs can be converted into LTM rings when $\alpha$ and $\alpha_{g}$ are simultaneously increased beyond $0.05$, a gain coefficient of such magnitude is physically impractical for realistic device implementation. In practical experimental scenarios, the achievable $\alpha_{g}$ typically falls within a moderate range of $0.01$-$0.03$ under feasible pumping. For example, as operation wavelength being $\lambda=1550$ $nm$, multi quantum well (MQW) consisting of InGaAsP can be electrically pumped to have $\alpha_{g}=-g_{mat}\lambda/(4\pi)=-0.0246$ with $g_{mat}=2000$ $cm^{-1}$ being the material gain \cite{gainMQW}. In contrast, both centrally inversion-symmetry-protected BICs and accidental BICs are highly susceptible to $\mathcal{PT}$-symmetric modulation and can be universally transformed into LTM rings for arbitrary valid $\alpha=\alpha_{g}$. This characteristic ensures that the $\mathcal{PT}$-symmetric meta-grating can consistently operate in the lasing regime, rendering the proposed physical phenomena experimentally observable with practical setups. Accordingly, this work adopts the parameter setting $\alpha=\alpha_{g}=0.05$ for subsequent investigations to comprehensively elucidate the underlying physics of lasing-assisted optical amplification and enhanced GH shifts. For realistic materials, the permittivity of the gain block should be dispersive, which can be described by Lorentz models \cite{lasingThreshold} given as $\varepsilon_{\alpha,g}=n_{r}^2-\alpha_{g}^{2}+\frac{2n_{r}\alpha_{g}\Gamma\omega_{1}}
{\omega_{1}^{2}-\omega^{2}+i\Gamma\omega}$ with $\omega_{1}$ being the center angular frequency, $\Gamma$ being the gain bandwidth, and $\omega=2\pi Re[f]$. By designing the structure of the MQW, $\omega_{1}$ is equal to the lasing frequency of the corresponding structure. As the frequency range of the quasi-BICs in the TE$_{3}$ band being smaller than the gain band width of $5\%$ $\omega_{1}$, the refractive index can be approximated as non-dispersive in the eigenvalue solver. Thus, we adopt the non-dispersive refractive index in the following discussing.

\section{GH shifts}

This section investigates the optical amplification of additionally incident weak optical beam irradiating on $\mathcal{PT}$-symmetric meta-gratings under stable lasing state. The incident beam intensity is assumed to be far lower than the saturation intensity of the gain medium. Under this weak-excitation approximation, the spatial profile of the saturated gain remains undisturbed, and the refractive index of the gain layer is fixed as $n_{g}=n_{r}-i\tilde{\alpha}_{g}(\mathbf{r})$, where $\tilde{\alpha}_{g}(\mathbf{r})$ denotes the saturated gain coefficient determined by the coexisting lasing field. A Gaussian beam with incident frequency $f_{inc}$, incident angle $\theta_{Inc}$, and beam width $w_{0}$ can be decomposed into a superposition of plane-wave components with identical frequency $f_{inc}$ and continuously varying incident angles $\theta_{in}$. The corresponding angular spectrum amplitude follows a Gaussian distribution: $\Theta(\theta_{in})=e^{-[(\theta_{in}-\theta_{Inc})/\Delta\theta_{Inc}]^{2}/2}$, where $\Delta\theta_{Inc}=\sin^{-1}[\lambda/(n_{b}\pi w_{0})]$ defines the beam divergence angle. For the resonant plane-wave component satisfying $f_{inc}=Re[f]$ and $\theta_{in}=\sin^{-1}[ck_{x}/(2\pi Re[f]n_{b})]\equiv\theta_{in,Res}$, the in-plane wave-vector matches the Bloch wave-vector of the resonant mode in the band structure, thereby enabling resonant mode excitation. As $\theta_{in}$ sweeps across the resonant angle $\theta_{in,Res}$, the complex reflection and transmission amplitudes, denoted as $r(\theta_{in})$ and $t(\theta_{in})$, undergo rapid angular variation.    The angular-dependent $r(\theta_{in})$ and $t(\theta_{in})$ are numerically computed via the finite element method within a periodic unit-cell domain. Upward and downward plane-wave incidences are implemented using periodic ports placed at $z=\pm(z_{max}-a)$, above and below the meta-grating, respectively. The reflection and transmission coefficients are extracted through Fourier integration of the scattered field $E_{y}^{sca}$ at the detection interfaces located at $z=\pm(z_{max}-a/2)$, formulated as $r=[1/(aE^{inc})]\int_{0}^{a}e^{ik_{x}x}E_{y}^{sca}(\mathbf{r})|_{z=s(z_{max}-a/2)}dx$ and $t=[1/(aE^{inc})]\int_{0}^{a}e^{ik_{x}x}E_{y}^{tot}(\mathbf{r})|_{z=-s(z_{max}-a/2)}dx$, with $s=\pm1$ representing incidence from upward and downward background, $E^{inc}$ representing incident amplitude, $E_{y}^{sca}(\mathbf{r})$ and $E_{y}^{tot}(\mathbf{r})$ representing the scattering and total fields, respectively. Finally, the full reflected and transmitted beam profiles are reconstructed by inverse Fourier transformation of the modulated angular spectra $\Theta(\theta_{in})r(\theta_{in})$ and $\Theta(\theta_{in})t(\theta_{in})$.

\subsection{Stationary-phase method}

When the beam width $w_{0}$ is considerably larger than the wavelength of the incident field, the incident Gaussian beam can be approximated as a quasi-plane wave with a vanishing divergence angle ($\Delta\theta_{Inc}\rightarrow0$). In this limit, the characteristics of reflected and transmitted beams are predominantly determined by the reflection and transmission coefficients evaluated at the central incident angle $\theta_{in}=\theta_{Inc}$. Based on the stationary-phase method, the lateral GH shifts for the reflected and transmitted beams are quantitatively defined as follows \cite{stationaryPt48}:
\begin{equation}
S_{GH,r/t}=-\frac{\lambda}{2\pi n_{b}\cos{\theta_{\mathrm{in}}}}\frac{\partial\phi_{r/t}}{\partial\theta_{\mathrm{in}}} \label{GH_SPM}
\end{equation}
where $\phi_{r/t}=\arg(r/t)$ denotes the phase of the corresponding reflection and transmission coefficients. In contrast, the angular GH shift describes the angular deviation of the output beams from the propagation angle predicted by Snell's law (i.e., $\theta_{Inc}$). The angular GH shifts for reflected and transmitted beams are formulated as \cite{angularGHshift}:
\begin{equation}
\Theta_{GH,r}=-\frac{\Delta\theta_{\mathrm{Inc}}^{2}}{2|r|}\frac{\partial|r|}{\partial\theta_{\mathrm{in}}}
\end{equation}
\begin{equation}
\Theta_{GH,t}=-\frac{\Delta\theta_{\mathrm{Inc}}^{2}}{2|t|}\frac{\partial|t|}{\partial\theta_{\mathrm{in}}}
\end{equation}
In conventional GH shift measurements, the incident angle is typically chosen to maximize the magnitude of the lateral GH shift, which coincides with the reflectance or transmittance peak. At such resonant peaks, the angular derivatives of the magnitude coefficients satisfy $\partial|r|/\partial\theta_{\mathrm{in}}=0$ and $\partial|t|/\partial\theta_{\mathrm{in}}=0$, such that the angular GH shift components can be safely neglected. Accordingly, the subsequent analysis focuses exclusively on the lateral GH shift.

\begin{figure}[tbp]
\scalebox{0.66}{\includegraphics{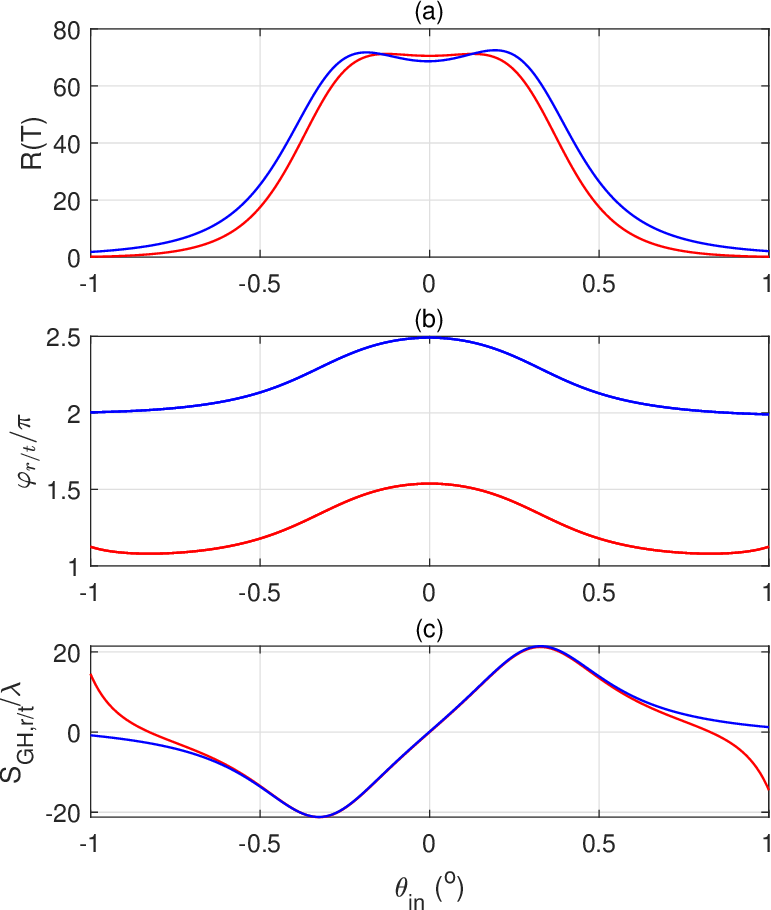}}
\caption{ For the Structure I bilayer meta-grating with $\alpha=\alpha_{g}=0.05$ under steady-state lasing conditions, panel (a) presents the angular spectra of reflectance (red curves) and transmittance (blue curves) for weak probe beams incident at $f_{inc}=0.447582c/a$. The corresponding phases of the reflection and transmission coefficients are illustrated in panel (b). Panel (c) shows the resultant GH shifts normalized by the incident wavelength, which are calculated via the stationary-phase method. }
\label{figure_gh755}
\end{figure}

Under the lasing state of Structure I, the resonant frequency of the dominant lasing mode is fixed at $f=0.447582c/a$. When the incident frequency satisfies $f_{inc}=f$, the angular spectra of reflectance and transmittance are plotted as the red and blue curves in Fig. \ref{figure_gh755}(a), respectively. Both the reflected and transmitted beam amplitudes achieve a nearly 70-fold amplification for incident angles smaller than $0.3^\circ$. The corresponding reflection and transmission phase spectra are presented in Fig. \ref{figure_gh755}(b), exhibiting remarkably steep phase slopes. Such drastic phase variation yields a prominent lateral GH shift with a maximum magnitude of $20\lambda$, as illustrated in Fig. \ref{figure_gh755}(c). Nevertheless, slight detuning of the incident frequency away from the resonant lasing frequency leads to a substantial decline in the optical amplification performance.

\begin{figure*}[tbp]
\scalebox{0.54}{\includegraphics{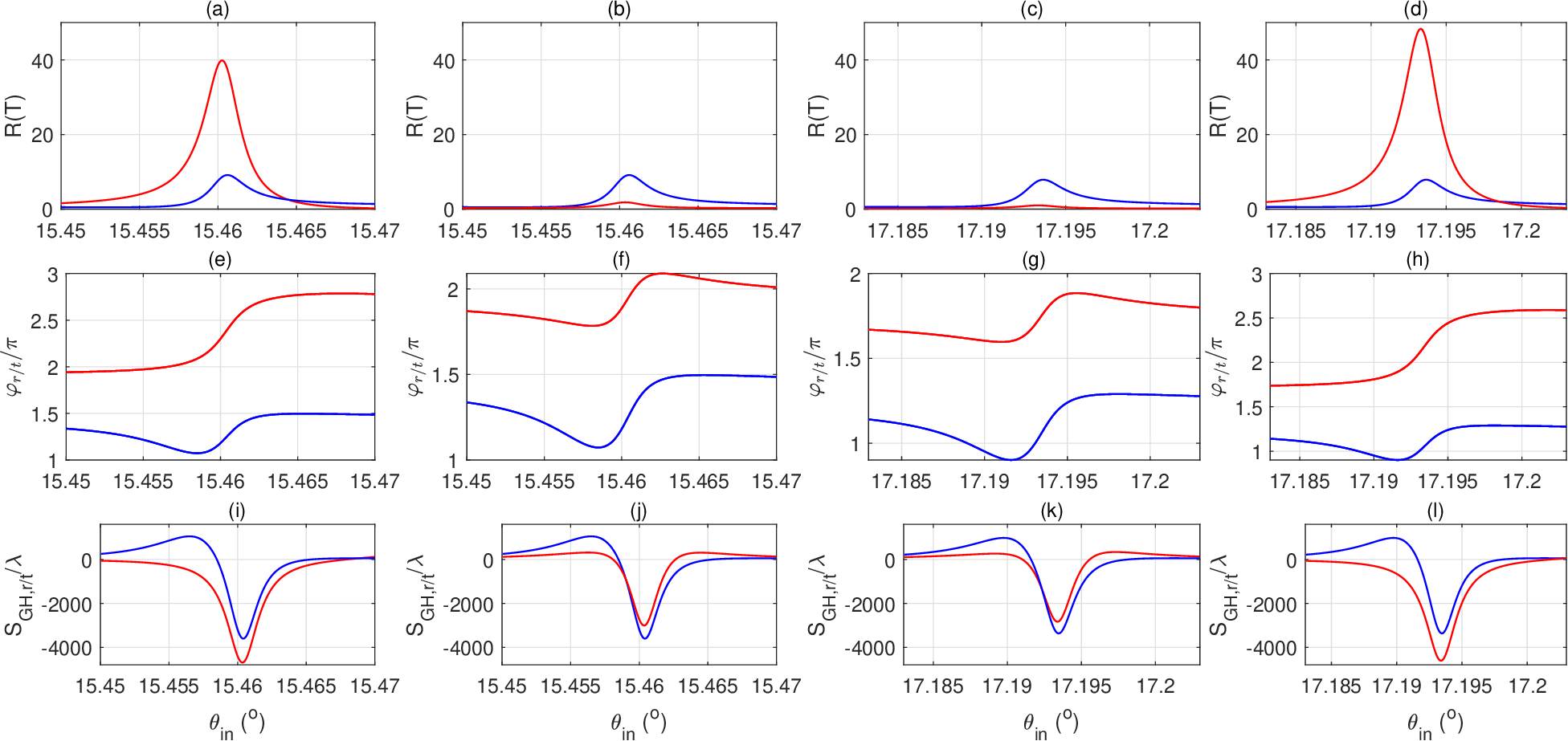}}
\caption{ For the Structure II bilayer meta-grating with $\alpha=\alpha_{g}=0.05$ under steady-state lasing conditions, the first row plots the angular spectra of reflectance (red curves) and transmittance (blue curves) for additionally incident weak optical fields. The second row presents the corresponding phases of the reflection and transmission coefficients, while the third row illustrates the resultant GH shifts normalized by the incident wavelength, calculated via the stationary-phase method. The first and third columns correspond to upward plane-wave incidence, whereas the second and fourth columns correspond to downward plane-wave incidence. Specifically, the optical fields in the first and second columns adopt an incident frequency of $f_{inc}=0.358462c/a$, associated with the resonant mode at $k_{x}=0.138(2\pi/a)$ in the band structure. By comparison, the third and fourth columns employ an incident frequency of $f_{inc}=0.356063c/a$, corresponding to the resonant mode at $k_{x}=0.152(2\pi/a)$.  }
\label{figure_gh1900}
\end{figure*}

\begin{figure}[tbp]
\centering
\scalebox{0.65}{\includegraphics{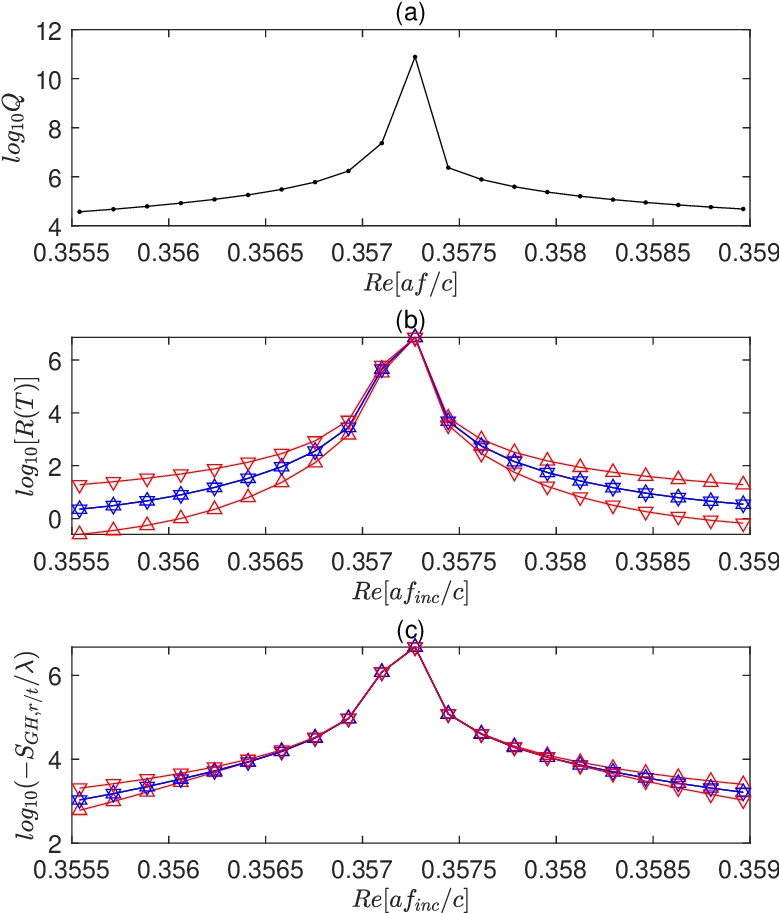}}
\caption{ For the Structure II bilayer meta-grating with $\alpha=\alpha_{g}=0.05$ under steady-state lasing conditions, panel (a) presents the logarithmic-scale Q factor of linear resonant modes in the TE$_{3}$ band as a function of the real resonant frequency. Panel (b) plots the logarithmic-scale peak values of reflectance (red dots) and transmittance (blue dots) extracted from the angular spectra of the additionally incident optical fields at different incident frequencies. The upward and downward quasi-plane-wave incidences are distinguished by upward- and downward-pointing triangular markers, respectively. Panel (c) shows the corresponding logarithmic-scale GH shifts normalized by the incident wavelength, which are evaluated at the resonant peaks in panel (b) and plotted using identical marker styles.  }
\label{figure_gh1900List}
\end{figure}

Under the stable lasing state of Structure II, the resonant frequency of the dominant lasing mode is fixed at $f=0.357271c/a$. When the incident frequency is slightly detuned above the lasing frequency, namely $f_{inc}=0.358462c/a$, nearby resonant modes satisfying $Re[f]=f_{inc}$ are effectively excited. These resonant modes exhibit negative group velocities, which, according to Eq. (\ref{GH_TCMT}), corresponds to negative GH shifts. The angular spectra of reflectance and transmittance for upward and downward plane-wave incidences are presented in Figs. \ref{figure_gh1900}(a) and (b), respectively. Both quantities far exceed unity, demonstrating prominent optical power amplification. The $\mathcal{PT}$ symmetry is well preserved under lasing state owing to the approximation of near threshold lasing with $\tilde{\alpha}_{g}(\mathbf{r})\approx\alpha_{g}$, which enforces identical transmittance responses for upward and downward incidence geometries.    The observed strong amplification originates from resonant mode excitation with suppressed radiative loss (small $Im[f]$) and enhanced radiative coupling strength (large $|c_{u/d}|$). In this near-resonant regime, the directionality factor $\eta$ approaches unity, yielding $|c_{u}|\gg|c_{d}|$. Since the mode-excitation amplitude is proportional to the radiative coupling coefficient $|c_{u/d}|$, upward incidence enables stronger resonant excitation and consequently produces larger reflectance, whereas downward incidence results in weaker reflection. The corresponding reflection and transmission phase spectra are plotted in Figs. \ref{figure_gh1900}(e) and (f). The resultant lateral GH shifts calculated via Eq. (\ref{GH_SPM}) are summarized in Figs. \ref{figure_gh1900}(i) and (j), confirming negative GH shifts at the resonant peaks with an ultra-large magnitude up to $4000\lambda$.

In contrast, when the incident frequency is slightly detuned below the lasing frequency ($f_{inc}=0.356063c/a$), the excited resonant modes still possess negative group velocities but feature a directionality factor close to $\eta=-1$. The resultant reflectance and transmittance spectra exhibit similar resonant amplification behaviors, with the primary difference being reversed directional asymmetry: downward incidence now generates significantly stronger reflection compared with upward incidence, as illustrated in Figs. \ref{figure_gh1900}(c) and (d). The associated angular spectra of phase are plotted in Figs. \ref{figure_gh1900}(g) and (h). The  corresponding lateral GH shifts are displayed in Figs. \ref{figure_gh1900}(k) and (l). At the resonant incident angles corresponding to the reflectance and transmittance peaks, the GH shifts remain negative and maintain a maximum magnitude of $4000\lambda$.

The spectral dependence of the resonant peak reflectance and transmittance over a broad frequency range covering the lasing frequency is summarized in Fig. \ref{figure_gh1900List}(b), while the corresponding resonant GH shift magnitudes are presented in Fig. \ref{figure_gh1900List}(c). The frequency-dependent Q factor of the resonant modes is plotted in Fig. \ref{figure_gh1900List}(a), which demonstrates that the Q factor rises asymptotically toward infinity as the resonant frequency approaches the lasing frequency. Comparative analysis of the numerical results reveals that both the optical amplification efficiency and the GH shift magnitude are positively correlated with the mode Q factor.

\subsection{Gaussian-beam incidence}

\begin{figure}[tbp]
\centering
\scalebox{0.66}{\includegraphics{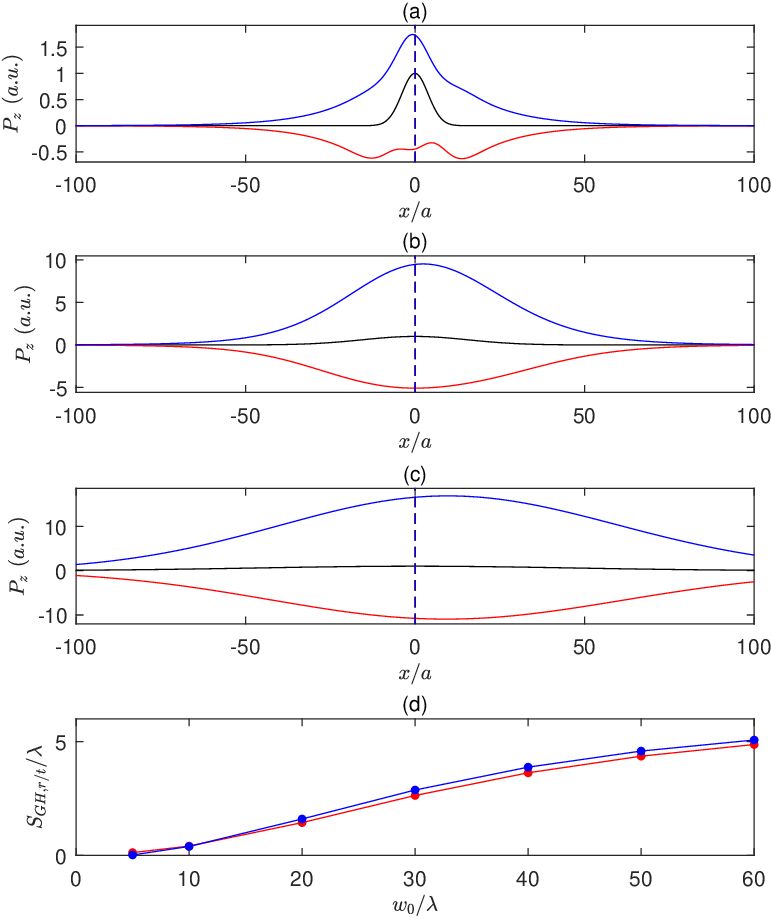}}
\caption{ For the finite-size Structure I bilayer meta-grating with $N_{x}=600$, $\alpha=\alpha_{g}=0.05$ under steady-state lasing conditions, panels (a)-(c) illustrate the transverse profiles of the $z$-component Poynting vector for downward-incident Gaussian beams with $\theta_{Inc}=0.29^\circ$ and $f_{inc}=0.447588c/a$. The black and red solid lines denote the incident and reflected field components at $z=-z_{max}$, while the blue solid lines represent the transmitted field component at $z=z_{max}$. All $P_{z}$ profiles are normalized such that the peak amplitude of the incident beam at $z=-z_{max}$ is unity. In panels (a)-(c), the geometrical beam centers of incident and reflected beams at $z=-z_{max}$ are plotted as vertical black and blue lines, respectively; those of transmitted beam at $z=z_{max}$ is at the blue line too. The Gaussian beam width is set as $w_{0}=5\lambda$, $20\lambda$, and $60\lambda$ in panels (a), (b), and (c), respectively. Panel (d) summarizes the numerically calculated GH shifts of the reflected and transmitted beams as functions of beam width, obtained from Eq. (\ref{numerGHshiftr}) and Eq. (\ref{numerGHshiftt}), which are plotted as dotted red and blue curves, respectively. }
\label{figure_slab755}
\end{figure}

In practical experimental implementations, meta-gratings possess finite lateral dimensions, typically consisting of  $600$ structural periods. The finite-size effect introduces additional radiative losses originating from structural truncation boundaries, which elevates the lasing threshold and reduces the field amplitude of the lasing mode. Accordingly, the spatial distribution of the saturated gain profile $\tilde{\alpha}_{g}(\mathbf{r})$ deviates from that of the ideal infinite structure. Furthermore, structural terminations reflect propagating resonant modes with opposite $k_{x}$ wave vectors, coupling counter-propagating modes into standing-wave field distributions. As a result, key lasing characteristics, including the lasing frequency, saturated gain landscape, and field patterns of all resonant modes (especially the lasing mode), are substantially modified by finite-size perturbations. The lasing behavior of finite meta-gratings can be analyzed following the same numerical procedure established in Section II. First, under the assumption of uniform gain distribution, the gain coefficient $\alpha_{g}$ is swept to determine the threshold value $\alpha_{g,th}$. At this threshold, only a single resonant mode emerges as the LTM with a purely real resonant frequency, while all other modes exhibit positive $Im[f]$ corresponding to radiative decay. This LTM is identified as the dominant lasing mode, and its time-averaged Poynting vector is substituted into Eq. (\ref{saturGain}) to derive the spatially dependent saturated gain profile. Subsequently, with the non-uniform saturated gain $\tilde{\alpha}_{g}(\mathbf{r})$ activated, the dimensionless lasing amplitude $I_{l}$ in Eq. (\ref{saturGain}) is finely tuned to acquire the steady-state lasing solution. The stable lasing condition is satisfied when exclusively the lasing mode maintains a real resonant frequency, whereas all other resonant modes retain positive imaginary frequency components. The far-field radiation pattern of the resonant mode under stable lasing states is then characterized to identify the optimal incident angle for resonant excitation by the additionally incident weak optical beam at its matched frequency, thereby achieving optical amplification and pronounced GH shifts.

Apart from structural finite-size effects, finite beam widths of practical Gaussian incident beams also suppress optical amplification and GH shifts. For a Gaussian beam with a finite waist $w_{0}$, its inherent divergence angle $\Delta\theta_{Inc}$ exceeds the narrow angular width of the resonant peaks in the reflectance and transmittance spectra. Consequently, only a fraction of the incident angular spectrum $\Theta(\theta_{\mathrm{in}})$ overlaps with the resonant region and contributes to the enhanced GH shift. As $w_{0}$ decreases, $\Delta\theta_{\mathrm{Inc}}$ increases correspondingly, further reducing the effective spectral overlap between the incident beam and the resonance. This leads to an overall reduction in the effective GH shift of the integrated reflected and transmitted beams \cite{fpcavityGH2,normalGH5}.

The combined influences of structural finite truncation and finite incident beam width jointly weaken the optical amplification and GH shift enhancement. The incident Gaussian beam simultaneously excite multiple resonant modes under the lasing state. The supperposition of the excited resonant modes determine the amplification and GH shift. To comprehensively characterize the beam-width-dependent GH shift suppression induced by the broadened incident angular spectrum and structural truncation, full-wave finite element simulations are performed for Gaussian beam scattering by finite-period meta-gratings. The numerical model adopts a computational domain spanning $|x|<x_{\mathrm{max}}=310a$ and $|z|<z_{max}=10a$, containing a finite meta-grating array of 600 periods located at $z=0$ within the region $|x|<300a$. Scattering boundary conditions with predefined Gaussian beam excitations are implemented at $z=+z_{max}$ and $z=-z_{max}$ in COMSOL Multiphysics to simulate upward and downward beam incidences, respectively. Given the highly localized nature of light-matter interaction near the meta-grating interface, the vertical separation between the scattering boundaries and the grating structure is sufficiently large to guarantee numerical stability and convergence.

For Structure I, the lasing mode resides at the $\Gamma$ point and possesses an extremely small lateral group velocity. Such characteristics render the lasing mode weakly susceptible to boundary reflection induced by structural truncation in finite-size meta-gratings. Consequently, finite-size effects barely alter the lasing threshold and the steady-state dimensionless lasing amplitude $I_{l}$. According to the numerical results in Fig. \ref{figure_gh755}(c), the GH shift reaches its maximum at an incident angle of $\theta_{\mathrm{in}}=0.3^\circ$. Accordingly, among all resonant modes of the finite metagrating under stable lasing state, the mode with $\mathrm{Re}[f]=0.447588c/a$ is selected for beam excitation, whose far-field radiation peak occurs at $0.29^\circ$ and thus closely matches the optimal resonant angle of $0.3^\circ$. This mode determines the incident Gaussian beam parameters adopted in subsequent simulations, with an incident frequency of $f_{\mathrm{inc}}=0.447588c/a$ and an incident angle of $\theta_{\mathrm{Inc}}=0.29^\circ$.
Under downward Gaussian beam incidence with a beam width of $w_{0}=5\lambda$, the vertical components of the Poynting vectors corresponding to the incident field $\hat{z}\cdot\mathbf{P}^{i}(\mathbf{r})|_{z=-z_{\mathrm{max}}}$, reflected field $\hat{z}\cdot\mathbf{P}^{s}(\mathbf{r})|_{z=-z_{\mathrm{max}}}$, and transmitted field $\hat{z}\cdot\mathbf{P}(\mathbf{r})|_{z=z_{\mathrm{max}}}$ are plotted as black, red, and blue curves in Fig. \ref{figure_slab755}(a), respectively. These curves characterize the transverse profiles of the incident, reflected, and transmitted beams at the corresponding observation planes, where $\mathbf{P}^{i}$, $\mathbf{P}^{s}$ and $\mathbf{P}$ denote the Poynting vectors of the incident, scattered, and total fields, respectively. The Gaussian beam is focused at the grating center $(x=0,z=0)$, such that the geometric beam center at the incidence plane $z=-z_{\mathrm{max}}$ is located at $x=-z_{\mathrm{max}}\sin\theta_{\mathrm{max}}$, as marked by the vertical dashed black line in Fig. \ref{figure_slab755}(a). Based on geometric optics, the theoretical beam centers of both reflected and transmitted beams at $z=-z_{\mathrm{max}}$ and $z=z_{\mathrm{max}}$ are positioned at $x=z_{\mathrm{max}}\sin\theta_{\mathrm{Inc}}$, corresponding to the vertical dashed blue line. Given the near-normal incidence condition ($\theta_{\mathrm{Inc}}\approx0$), the two dashed lines nearly overlap at $x=0$.
To quantitatively evaluate the actual beam shift, the numerical GH shift of the reflected beam is defined as the energy-flux-weighted lateral displacement relative to the geometric optical position, formulated as:
\begin{equation}
S_{GH,r}=\frac{-\int_{-x_{max}}^{x_{max}}{x\hat{z}\cdot\mathbf{P}^{s}(\mathbf{r})|_{z=-10a}dx}}{-\int_{-x_{max}}^{x_{max}}{\hat{z}\cdot\mathbf{P}^{s}(\mathbf{r})|_{z=-10a}dx}}-z_{max}\sin\theta_{Inc}
\label{numerGHshiftr}
\end{equation}
The weight is normalized by the total reflected power given as $-\int_{-x_{max}}^{x_{max}}{\hat{z}\cdot\mathbf{P}^{s}(\mathbf{r})|_{z=-10a}dx}$.
Similarly, the numerical GH shift of the transmitted beam is calculated via:
\begin{equation}
S_{GH,t}=\frac{\int_{-x_{max}}^{x_{max}}{x\hat{z}\cdot\mathbf{P}(\mathbf{r})|_{z=10a}dx}}{\int_{-x_{max}}^{x_{max}}{\hat{z}\cdot\mathbf{P}(\mathbf{r})|_{z=10a}dx}}-z_{max}\sin\theta_{Inc}
\label{numerGHshiftt}
\end{equation}
For the tightly focused incident beam with $w_{0}=5\lambda$, both reflected and transmitted beams exhibit obvious field amplification but negligible GH shift, with calculated $S_{GH,r}$ and $S_{GH,t}$ values far smaller than $\lambda$. In addition, uneven amplification of individual angular spectral components distorts the beam profiles, resulting in non-Gaussian field distributions. In contrast, for wider incident beams with $w_{0}=20\lambda$ and $w_{0}=60\lambda$, the reflected and transmitted beam profiles displayed in Figs. \ref{figure_slab755}(b) and (c) demonstrate enhanced optical amplification and prominent finite GH shifts. Meanwhile, the output beam profiles recover nearly ideal Gaussian shapes due to improved spectral uniformity. The extracted GH shift values for varying beam widthes are summarized in Fig. \ref{figure_slab755}(d), showing that the GH shift asymptotically converges to $5\lambda$ as $w_{0}$ increases. Overall, the proposed $\mathcal{PT}$-symmetric metagrating enables simultaneous realization of substantial optical amplification and large GH shift under near-normal Gaussian beam incidence.

\begin{figure}[tbp]
\centering
\scalebox{0.66}{\includegraphics{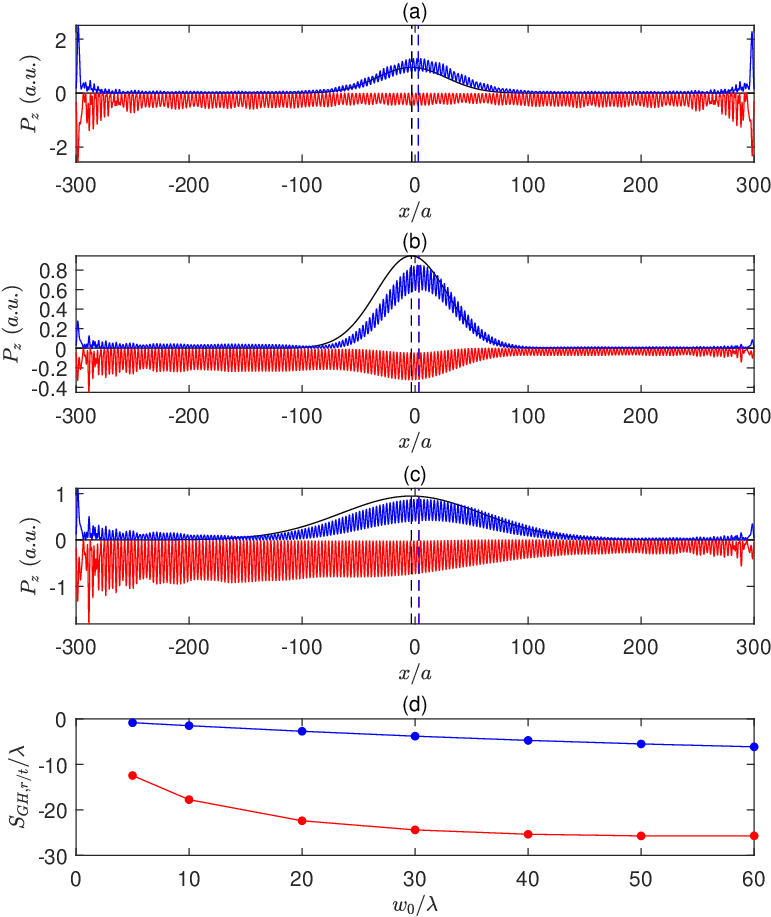}}
\caption{ This figure presents plots obtained for Structure II, following the same plotting conventions as Fig. \ref{figure_slab755}. In panel (a), $f_{inc}=0.356052c/a$, $\theta_{Inc}=17.21^\circ$, and $w_{0}=30\lambda$. In panels (b)-(d), $f_{inc}=0.353543c/a$ and $\theta_{Inc}=19.02^\circ$. $w_{0}=30\lambda$ and $w_{0}=60\lambda$ are adopted in panels (b) and (c), respectively. Panel (d) summarizes the numerically calculated GH shifts of the reflected and transmitted beams as functions of beam width, obtained from Eq. (\ref{numerGHshiftr}) and Eq. (\ref{numerGHshiftt}), which are plotted as dotted red and blue curves, respectively. }
\label{figure_slab1900}
\end{figure}

Unlike Structure I, the lasing mode of Structure II locates away from the $\Gamma$ point and possesses a considerable lateral group velocity. This feature renders the mode highly susceptible to boundary reflection at the truncated edges of finite-size meta-gratings. Accordingly, finite-size effects profoundly modify the lasing threshold and the dimensionless steady-state lasing amplitude $I_{l}$. Numerically, the saturated lasing amplitude of the finite meta-grating is calculated as $0.0013$, which is notably lower than the value of $0.003742025$ obtained for the ideal infinite meta-grating in Section II. Under stable lasing conditions, resonant modes in the spectral vicinity of the lasing mode evolve into standing-wave modes, composed of counter-propagating Bloch waves with nearly equal amplitudes. Such standing-wave formation strongly suppresses the GH shift, representing a dominant finite-size-induced limitation.
For the resonant mode with $\mathrm{Re}[f]=0.356052c/a$[the closest to the incident frequency in Figs. \ref{figure_gh1900}(c,d,g,h,k,l)], the far-field radiation maximum occurs at an angle of $17.21^\circ$. A Gaussian beam with identical frequency, incident angle, and a beam width of $w_{0}=30\lambda$ is adopted for excitation. The resultant reflected and transmitted beam profiles are plotted in red and blue in Fig. \ref{figure_slab1900}(a), respectively. The reflected beam exhibits a flattened, non-Gaussian profile characterized by periodic interference fringes with a uniform period of approximately $3.3a$. These fringes originate from the superposition of counter-propagating Bloch waves with $k_{x}=\pm0.152(2\pi/a)$, yielding a wavevector difference of $0.304(2\pi/a)$ and verifying the formation of standing-wave field distributions. The symmetric field magnitude distribution on both sides of the beam center results in a nearly vanishing GH shift.
In contrast, for incident frequencies further detuned from the lasing frequency, the theoretical GH shift of the infinite meta-grating decreases significantly, thereby weakening the relative impact of finite-size perturbations. A representative example is demonstrated in Fig. \ref{figure_slab1900}(b), where the Gaussian beam parameters are adjusted to $f_{\mathrm{inc}}=0.353543c/a$ and $\theta_{\mathrm{Inc}}=19.02^\circ$, corresponding to the resonant Bloch mode at $k_{x}=\pm0.166(2\pi/a)$ in the infinite structure. The reflected beam profile exhibits asymmetric field distribution, with enhanced amplitude on the left side of the beam center, which corresponds to a negative GH shift of $-12\lambda$. As the beam width increases to $w_{0}=60\lambda$, the reflection amplification is further improved, as depicted in Fig. \ref{figure_slab1900}(c). Sweeping the beam width from $5\lambda$ to $60\lambda$ reveals that the numerical GH shift gradually converges to $-26\lambda$ (Fig. \ref{figure_slab1900}(d)). For comparison, the ideal infinite meta-grating achieves a much larger negative GH shift of $-366\lambda$ at the optimal incident angle of $18.97^\circ$ under the same incident frequency. These results indicate that the theoretical large GH shift can be asymptotically recovered in finite meta-gratings with sufficiently large beam width of the incident Gaussian beams.

\section{Conclusion}

In summary, the coexisting gain and loss in $\mathcal{PT}$-symmetric bilayer meta-gratings enable BIC-hosting systems to operate slightly above the lasing threshold. Under stable lasing state, a single resonant mode selected from the continuous LTM ring serves as the dominant lasing mode, which originates from either central-symmetry-protected BICs or accidental BICs in the Hermitian limit. Upon additional illumination by a weak optical beam with negligible perturbation to the stabilized lasing background, the angular spectra of the scattered fields exhibit pronounced resonant peaks accompanied by simultaneous optical amplification and enhanced GH shifts. For configurations where the lasing mode evolves from a $\Gamma$-point central-symmetry-protected BIC, both the field amplification and GH shift remain prominent. In practical finite meta-gratings, the finite size effect hardly influence the lasing state, the amplification or the GH shift of the additionally incident optical beam. In contrast, for configurations where the lasing mode evolves from  accidental BICs off $\Gamma$ point, the magnitudes of the amplification and GH shift are positively correlated with the Q factor of the resonant mode under lasing state, which asymptotically diverges as the frequency approaches the lasing frequency. In practical finite meta-gratings, the finite size effect slightly raise the lasing threshold and suppress the lasing mode, as well as greatly restrict the amplification and GH shift of the additionally incident optical beam.

\begin{acknowledgments}
This project is supported by the Special Projects in Key Fields of Ordinary Universities in Guangdong Province(New Generation Information Technology, Grant No. 2023ZDZX1007) and the Natural Science Foundation of Guangdong Province of China (Grant No. 2026A1515012428).
\end{acknowledgments}


\clearpage

\end{document}